\documentclass[aps, prl, reprint, superscriptaddress]{revtex4-2}
\usepackage{graphicx, subfigure}
\usepackage{amsmath,amssymb,epstopdf}
\usepackage{color}
\usepackage{natbib,hyperref}
\usepackage{multirow}
\usepackage{pdfpages}

\makeatletter
\AtBeginDocument{\let\LS@rot\@undefined}
\makeatother

\begin{document}

\title{Microscopic bath effects on noise spectra in semiconductor quantum dot qubits}

\author{Seongjin Ahn}
\author{S.~Das Sarma}
\affiliation{Condensed Matter Theory Center and Joint Quantum Institute, Department of Physics, University of Maryland College Park, MD 20742, USA}
\author{J.~P.~Kestner}
\email{jkestner@umbc.edu}
\affiliation{Department of Physics, University of Maryland Baltimore County, Baltimore, MD 21250, USA}

\begin{abstract}
When a system is thermally coupled to only a small part of a larger bath, statistical fluctuations of the temperature (more precisely, the internal energy) of this ``sub-bath" around the mean temperature defined by the larger bath can become significant.  We show that these temperature fluctuations generally give rise to 1/f-like noise power spectral density from even a single two-level system.  We extend these results to a distribution of fluctuators, finding the corresponding modification to the Dutta-Horn relation.  Then we consider the specific situation of charge noise in silicon quantum dot qubits and show that recent experimental data [E. J. Connors, {\it et al}., Phys. Rev. B \textbf{100}, 165305 (2019)] can be modeled as arising from as few as two two-level fluctuators, and accounting for sub-bath size improves the quality of the fit.
\end{abstract}

\maketitle
Charge noise, particularly the so-called 1/f noise ubiquitous in electronic devices, is currently the most significant roadblock to the successful development of semiconductor-based scalable solid state qubits. It is well known that an ensemble of thermally activated two-level fluctuators (TLFs) with a broad range of switching rates gives rise to a 1/f power spectral density (PSD) with a linear temperature dependence \cite{Kogan1996}.  This is the standard hand-waving explanation given to explain  wide-spread observations of pink noise in solid state qubit devices, with some sort of charged defects playing the role of the TLFs.  Early data from laterally-defined quantum dots in silicon showed that the noise power indeed appears to increase with temperature \cite{Freeman2016}, but with large error bars that preclude a more detailed conclusion.  Recent experiments, however, have shown striking deviation from the expected linear temperature dependence. Ref.~\cite{Connors_2019} measures a temperature dependence which is not only nonlinear, but in some cases \emph{nonmonotonic}, qualitatively consistent with the Dutta-Horn model \cite{Dutta1979} of a large ensemble of TLFs with a nonuniform distribution of switching rates, although any expected quantitative consistency varies widely within the data set. Meanwhile, Ref.~\cite{Petit_2018} finds a quadratic temperature dependence.  On the other hand, Ref.~\cite{Petit_2020} observes a $T_2^{\ast}$ decoherence time that is approximately \emph{constant} over a range of temperatures from 0.45K-1.2K, suggesting that their charge noise comes from a few TLFs with activation energies much smaller than $k_B T$, rather than a broad distribution. The situation, particularly in the measured noise temperature dependence, is thus quite confusing.

In this work we show that, in principle, 1/f noise with nonlinear temperature dependence can be produced by even a single TLF coupled to a microscopic subsection of the thermal bath.  Although we cannot assert that our proposed mechanism is operational in semiconductor qubits (in fact, the physical mechanism underlying 1/f noise is still obscure), we show that the experimental data of Ref.~\cite{Connors_2019} can be reasonably fit as arising from a small number of TLFs, and that the fit is improved by incorporating this microscopic thermal bath effect via an additional fit parameter.

The essential narrative here is that, even within the TLF model, observation of 1/f noise over some broad frequency range need not imply an ensemble of TLFs.  One (or few) TLFs can suffice, in which case a nonlinear temperature dependence is natural. This conclusion of the adequacy of just a few TLFs obviously has important implications.

It is quite physical to assume that a microscopic two-level fluctuator may be coupled to only a microscopic part of the thermal bath.  For instance, in semiconductor quantum dots one typically imagines the charge noise as arising from charged TLFs coupled via Coulomb interaction to a thermal electrostatic environment, but since the Coulomb interaction is strongly screened by the nearby two-dimensional electron gas (2DEG) and the dense array of metallic top gates, each TLF will only interact with a region of the bath that lies less than a screening length away.  This ``sub-bath" has an internal energy characterized by an effective temperature, $T_{sb}$, whose average is the same as the macroscopic bath temperature, $T$, but with fluctuations around equilibrium of variance
\begin{equation}\label{eq:sigmasb1}
\sigma_{sb}^2 = \frac{k_B T^2}{C_V},
\end{equation}
where $C_V$ is the heat capacity of the sub-bath.  Since $C_V$ is an extrinsic quantity, $\sigma_{sb}$ is proportional to one over the square root of the volume of the sub-bath, which becomes significant at small enough volumes.  Generally, $C_V$ is also a function of $T$; at low temperatures the phonon contribution is negligible and it is dominated by the electronic heat capacity, which is linear in $T$.  For example, considering a 2DEG sub-bath of area $A$, the variance in temperature $T_{sb}$ is
\begin{equation}\label{eq:sigmasb2}
\sigma_{sb}^2 = \frac{3 \hbar^2 T}{\pi m A k_B},
\end{equation}
where $m$ is the effective mass of the electrons.  This can certainly be non-negligible, since for Si, with an effective mass of $m=0.19 m_e$ (where $m_e$ is the electron mass), and an area of one square micron, at a typical temperature of 50 mK one would have sub-bath fluctuations of 14 mK.

We thus assume a stochastic TLF with activation energy $E$ and a thermally activated transition time $\tau \exp\left(E/k_B T_{sb}\right)$.  Averaging over the sub-bath statistical temperature distribution $f(T_{sb})$, the PSD is
\begin{multline}\label{eq:S1}
  S(\omega)= \Delta^2 \int_{0}^{\infty} dT_{sb} f(T_{sb}) \frac{4 \tau \exp\left(E/k_B T_{sb}\right)}{1+\omega^2\tau^2\exp\left(2 E/k_B T_{sb}\right)}
  \\
  =\frac{2 \Delta^2}{\omega}\int_{0}^{\infty} dT_{sb} f(T_{sb}) \text{sech} \left(\frac{E}{k_B T_{sb}} + \ln (\omega\tau)\right),
\end{multline}
where $\Delta^2$ is the total variance of the signal produced by the switching events.
Integrating over a gaussian distribution of temperatures, truncated to avoid unphysical negative temperatures and normalized,
\begin{equation}\label{eq:S2}
S(\omega) = \frac{2 \Delta^2}{\omega}\int_{0}^{\infty} dT_{sb} \frac{e^{-\frac{\left(T_{sb}-T\right)^2}{2\sigma_{sb}^2}} \text{sech}\left(\frac{E}{k_B}\left(\frac{1}{T_{sb}} - \frac{1}{T_{\omega}}\right)\right)}{\sqrt{\pi/2} \sigma_{sb}\left(1+\text{erf}\left(\frac{T}{\sqrt{2} \sigma_{sb}}\right)\right)},
\end{equation}
where we have also defined
\begin{equation}\label{eq:Tw}
T_{\omega}\equiv \frac{E}{k_B \ln \frac{1}{\omega\tau}}.
\end{equation}

The integrand in Eq.~\eqref{eq:S2} contains two peaks, one at the distribution's mean, $T$, and one from the Lorentzian at $T_{\omega}$.  While the integral is easily carried out numerically for any set of parameters, the qualitative behavior is illuminated by the following approximation. Assuming the peaks are well separated, approximate the sech term as a constant in the vicinity of $T$ and as a gaussian in the vicinity of $T_{\omega}$, with width
\begin{equation}\label{eq:sigmaw}
  \sigma_{\omega} \equiv \frac{E}{k_B \ln^2 \frac{1}{\omega\tau}},
\end{equation}
so as to obtain
\begin{multline}\label{eq:S3}
S(\omega) \approx \frac{2\Delta^2}{\omega}\text{sech}\left(\frac{E}{k_B}\left(\frac{1}{T} - \frac{1}{T_{\omega}}\right)\right)
\\
+\frac{2\Delta^2}{\omega} \frac{\sigma_{\omega}}{\sqrt{\sigma_{sb}^2+\sigma_{\omega}^2}} \frac{1+\text{erf}\left(\frac{\sigma_{sb}^2 T_{\omega} + \sigma_{\omega}^2 T}{\sqrt{2}\sigma_{sb}\sigma_{\omega}\sqrt{\sigma_{sb}^2+\sigma_{\omega}^2}}\right)}{1+\text{erf}\left(\frac{T}{\sqrt{2}\sigma_{sb}}\right)} e^{-\frac{\left(T-T_{\omega}\right)^2}{2(\sigma_{sb}^2+\sigma_{\omega}^2)}}.
\end{multline}

\begin{figure}
  \centering
  \includegraphics[width=\columnwidth]{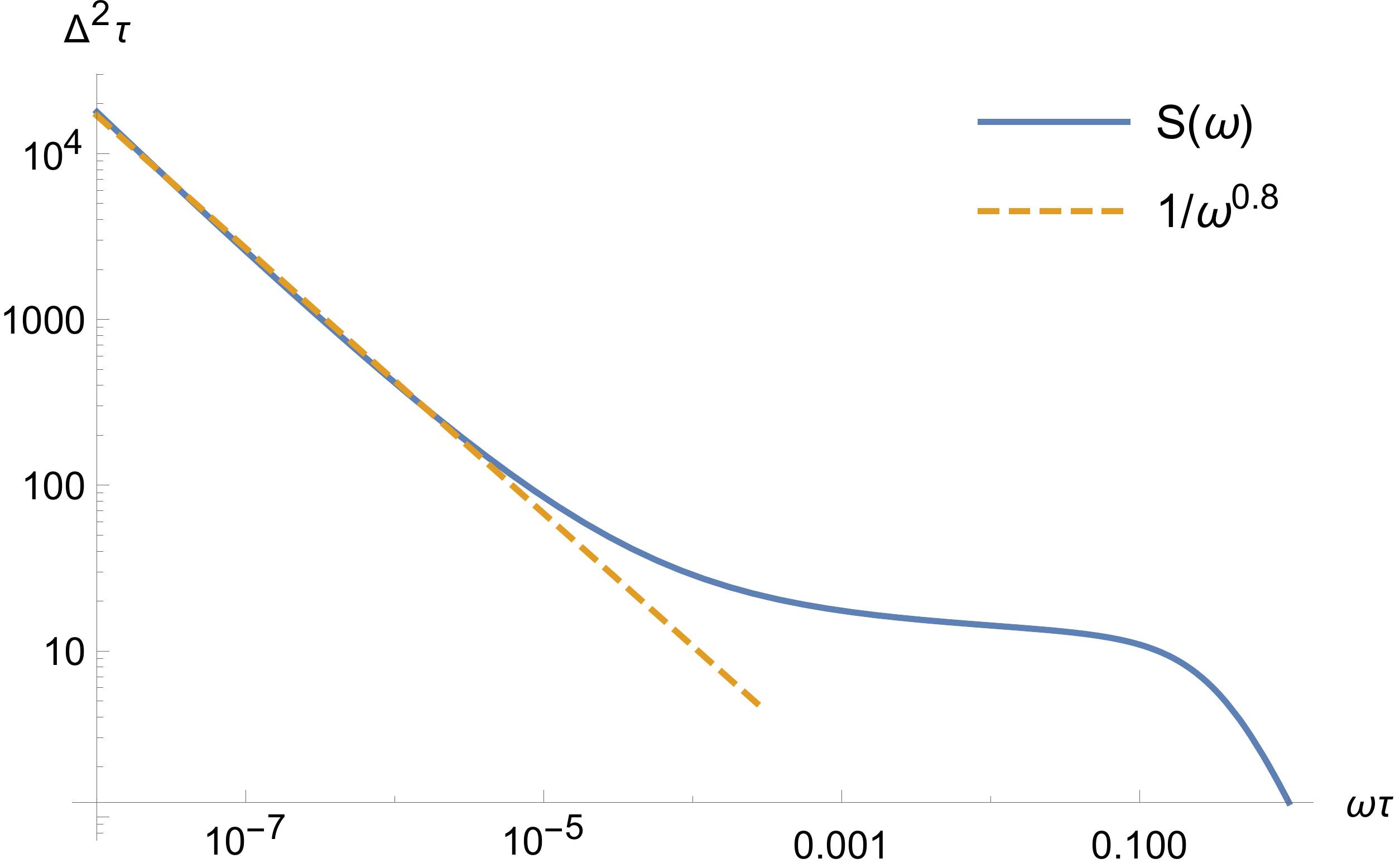}
  \caption{Power spectral density vs frequency from Eq.~\eqref{eq:S2} for $\frac{E}{k_B T} = 1$, $\frac{\sigma_{sb}}{T} = 0.3$.}\label{fig:S}
\end{figure}

This has nontrivial frequency and temperature dependence, but in the low-frequency limit $T_{\omega} \ll T$ and $\sigma_{\omega} \ll \sigma_{sb}$,
\begin{equation}\label{eq:Sapprox}
S(\omega) \approx \frac{4\tau\Delta^2}{1+\omega^2\tau^2} + \frac{2E\Delta^2}{k_B \sigma_{sb} \omega \ln^2 \frac{1}{\omega\tau}} \frac{2 e^{-\frac{\left(T-T_{\omega}\right)^2}{2\sigma_{sb}^2}}}{1+\text{erf}\left(\frac{T}{\sqrt{2}\sigma_{sb}}\right)}.
\end{equation}
The first term is the typical Lorentzian spectrum of a single TLF. The second term rises above this Lorentzian floor at low frequencies as $1/\omega \ln^2 \omega\tau$, which over a broad frequency range centered around $\omega_0$ is practically indistinguishable from $1/\omega^{\alpha}$ with $\alpha = 1+2/\ln\omega_0\tau$.  An example of this behavior is shown in Fig.~\ref{fig:S}.

Thus, if temperature fluctuations are significant, one can expect to find $1/\omega^{\alpha}$ noise at frequencies below roughly
\begin{equation}\label{eq:wcrit}
  \omega_{\text{1/f}} \sim 0.01\tau^{-1} \frac{E}{k_B \sigma_{sb}\left(1+\text{erf}\left(\frac{T}{\sqrt{2}\sigma_{sb}}\right)\right)}\exp\left(-\frac{T^2}{2\sigma_{sb}^2}\right),
\end{equation}
transitioning to white noise at intermediate frequencies, and finally falling as $1/\omega^2$ at frequencies above $\tau^{-1}$.  Although no quantum dot experiment to our knowledge has measured the noise spectrum at high enough frequency to conclusively observe the roll-off to $1/\omega^2$ (although Ref.~\cite{Gungordu2019} finds suggestions of it in the data of Ref.~\cite{Yoneda_2017}), some have observed a whitening of $1/\omega^{\alpha}$ noise with increasing frequency \cite{Connors_2019,Chan2018}.

\begin{figure}
  \centering
  \includegraphics[width=\columnwidth]{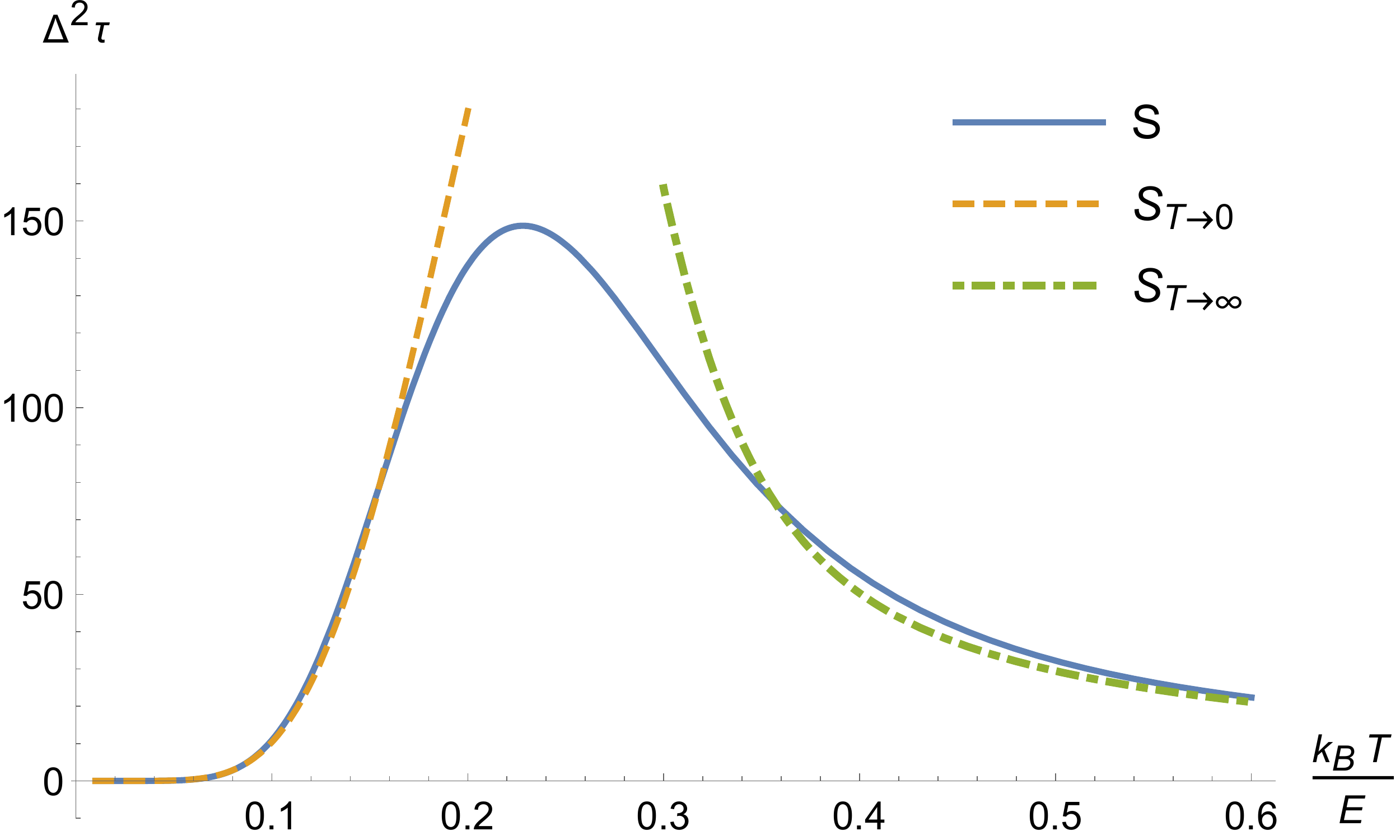}
  \caption{Power spectral density at $\omega\tau = 0.01$ vs temperature for $\frac{k_B \sigma_{sb}}{E} = 0.1\sqrt{k_B T/E}$.  The solid line is numerically computed from from Eq.~\eqref{eq:S2}, the dashed line is from Eq.~\eqref{eq:SlowT2}, and the dotted line is from Eq.~\eqref{eq:Sapprox}.}\label{fig:SvsT}
\end{figure}

Turning our attention now to the temperature dependence, we can compute Eq.~\eqref{eq:S2} numerically for a given set of parameters, as shown in Fig.~\ref{fig:SvsT}.
The PSD is peaked around $T_\omega$ and is qualitatively similar to the PSD in the absence of temperature fluctuations.
It is instructive to look at the limiting cases analytically.  At temperatures well above $T_{\omega}$, Eq.~\eqref{eq:Sapprox} holds, so one has an exponential decay to a constant value as temperature increases.  This is the same high-temperature dependence as in the absence of fluctuations.  The data in Ref.~\cite{Petit_2020} is believed to correspond to this constant, high-temperature tail.  For $T < T_\omega$ (and hence also $< E/k_B$), the peaks of the integrand in Eq.~\eqref{eq:S2} at $T$ and $T_{\omega}$ are no longer well separated. However, by approximating the sech as a decaying exponential (i.e., neglecting unity in the denominator of Eq.~\eqref{eq:S1} compared to the exponentially large term), we can make a gaussian approximation to the integrand
\begin{multline}\label{eq:SlowT1}
S(\omega)\overset{T\rightarrow 0}{\approx} \frac{4\Delta^2}{\omega^2\tau} \int_{0}^{\infty} dT_{sb} \frac{e^{-\frac{E}{k_B T_{sb}}-\frac{\left(T_{sb}-T\right)^2}{2\sigma_{sb}^2}}}{\sqrt{\pi/2} \sigma_{sb} \left(1+\text{erf}\left(\frac{T}{\sqrt{2}\sigma_{sb}}\right)\right)}
\\
\approx \frac{4\Delta^2}{\omega^2\tau} \frac{e^{-\frac{E}{k_B T_{\ast}}-\frac{\left(T_{\ast}-T\right)^2}{2\sigma_{\ast}^2}}}{\sqrt{\pi/2} \sigma_{sb} \left(1+\text{erf}\left(\frac{T}{\sqrt{2}\sigma_{sb}}\right)\right)}\int_{0}^{\infty} dT_{sb} e^{-\frac{\left(T_{sb}-T_{\ast}\right)^2}{2\sigma_{\ast}^2}},
\end{multline}
where the expressions for $T_{\ast}$ and $\sigma_{\ast}$ in terms of the physical parameters are algebraically cumbersome, but simplify in the low-temperature limit if we assume $\lim_{T\rightarrow 0} T/\left(\sigma_{sb}^2 E/k_B\right)^{1/3} = 0$ (as is the case for electrons \eqref{eq:sigmasb2}) to
\begin{equation}\label{eq:Tast}
T_{\ast} = \frac{1}{3}T+\left(\frac{E}{k_B}\sigma_{sb}^2\right)^{1/3}, \qquad \sigma_{\ast} = \frac{1}{\sqrt{3}}\sigma_{sb},
\end{equation}
and one obtains
\begin{multline}\label{eq:SlowT2}
S(\omega)\overset{T\rightarrow 0}{\approx} \frac{4\sqrt{2}\Delta^2}{\sqrt{3}\omega^2\tau}\exp\left(-\frac{1}{2}\left(-\frac{2T}{3\sigma_{sb}} + \left(\frac{E}{k_B \sigma_{sb}}\right)^{1/3}\right)^2\right)
\\
\times\exp\left(-\frac{3E/k_B}{T+3\left(\sigma_{sb}^2 E/k_B\right)^{1/3}}\right).
\end{multline}
So, instead of the typical linear temperature dependence \cite{Kogan1996}, it is exponentially flat at low temperatures, going like $\exp\left(-\left(\frac{3\sqrt{\pi m A} E}{2\sqrt{2 k_B} \hbar}\right)^{2/3}T^{-1/3}\right)$.

This flatness at low temperatures is qualitatively consistent with some of the data sets of Ref.~\cite{Connors_2019}, and we show below that superposing PSDs as in Fig.~\ref{fig:SvsT} of a few TLFs with different activation energies can provide an alternate way to understand and fit the nonlinear behavior in Ref.~\cite{Connors_2019}.  However, at this point it suffices to note that by dropping the assumption that 1/f noise must imply a continuous distribution of TLFs, we have preserved the nonlinear temperature dependence of a single TLF, and, depending on the parameter values, one could obtain seemingly very different behaviors if one observes over only a narrow range of temperatures.

We now discuss briefly how the PSD of a continuous distribution of TLFs would be affected by sub-bath temperature fluctuations.  The classic Dutta-Horn model \cite{Dutta1979} for a temperature-independent distribution of activation energies, $F(E)$, gives
\begin{equation}\label{eq:DHS}
S = \frac{2\pi k_B T}{\omega} F\left(E_{\omega}\right),
\end{equation}
where $E_{\omega}\equiv k_B T \ln \frac{1}{\omega \tau}$, in which case the frequency dependence and the temperature dependence are linked as
\begin{equation}\label{eq:DHgamma}
\gamma \equiv -\frac{\partial \ln S}{\partial \ln \omega} = 1 - \frac{1}{\ln \omega \tau}\left(\frac{\partial \ln S}{\partial \ln T}-1\right).
\end{equation}

Modifying Eq.~\eqref{eq:DHS} to the case of a fluctuating sub-bath temperature [more precisely, starting from Eq.~\eqref{eq:S1} and assuming $F$ is broad, with a width much larger than $k_B T$ such that the sech function can be approximated as a delta function],
\begin{equation}\label{eq:DHSwithTsb}
S = \frac{2\pi k_B\Delta^2}{\omega} \int dT_{sb} T_{sb} F(kT_{sb} \ln \frac{1}{\omega \tau})f\left(T_{sb}\right).
\end{equation}
Since the distribution of activation energies is assumed narrow compared to the distribution of sub-bath temperatures, we can do the integration in Eq.~\eqref{eq:DHSwithTsb} by approximating
\begin{multline}
T_{sb} F(k_B T_{sb} \ln \frac{1}{\omega \tau}) \approx T F\left(E_{\omega}\right)
\\
+\left( F\left(E_{\omega}\right) + E_{\omega}F'\left(E_{\omega}\right) \right)\left(T_{sb}-T\right)
\\
+\left(\frac{E_{\omega}}{T}F'\left(E_{\omega}\right) + \frac{E_{\omega}^2}{2 T}F''\left(E_{\omega}\right)\right)\left(T_{sb}-T\right)^2.
\end{multline}
Plugging this into Eq.~\eqref{eq:DHSwithTsb}, and assuming for simplicity that $\sigma_{sb} < T$ such that we can neglect the truncation of the gaussian temperature distribution and allow the tail to extend slightly into unphysical negative sub-bath temperatures, we obtain
\begin{multline}
    S(\omega)=\frac{2\pi k_B T\Delta^2}{\omega}F\left(E_{\omega}\right)
             \\
             + \frac{\pi k_B^2\sigma_{sb}^2}{\omega}\left[2\ln \frac{1}{\omega \tau} F'\left(E_{\omega}\right) + E_{\omega} \ln \frac{1}{\omega \tau}  F''\left(E_{\omega}\right)\right].
\end{multline}
(One can easily do the integral with the truncated gaussian distribution instead, picking up additional terms with factors of $e^{-T^2/\sigma_{sb}^2}$, but it does not affect the conclusions and the results are algebraically unwieldy, so we do not reproduce them here.)

Then it is straightforward to obtain the corresponding approximate modified Dutta-Horn relationship:
\begin{widetext}
\begin{equation}
    \gamma=1-\frac{1}{\ln \omega\tau}\left(\frac{\partial\ln{S}}{\partial\ln{T}}-1\right)\left(1+\frac{2\sigma_{sb}^2\left(2 F'\left(E_{\omega}\right) + E_{\omega} F''\left(E_{\omega}\right)\right)}{2\left(T^2-\sigma_{sb}^2\right) F'\left(E_{\omega}\right) + E_{\omega} \sigma_{sb}^2 \left(2 F''\left(E_{\omega}\right) + E_{\omega} F'''\left(E_{\omega}\right)\right)}\right).
\end{equation}
\end{widetext}
The main point here is that including temperature fluctuations destroys the key feature of the Dutta-Horn result that the relationship between the frequency dependence and temperature dependence is independent of the details of the activation energy distribution.  Only in the restricted case of negligible second- and higher-order derivatives of $F$ does the relationship becomes independent of the form of $F$:
\begin{equation}
    \gamma=1-\frac{1}{\ln{(\omega\tau)}}\left(\frac{\partial\ln{S}}{\partial\ln{T}}-1\right)
    \frac{1+\sigma_{sb}^2/T^2}
         {1-\sigma_{sb}^2/T^2}.
\end{equation}
Generally, the frequency and temperature dependences are now decoupled in the sense that one cannot predict one from the other without knowing the underlying distribution. In such a scenario, the details of the noise would matter a great deal, leading possibly to nonuniversal experimental behavior.

We now turn to the experimental data of Reference \cite{Connors_2019}.  There the charge noise in several silicon double quantum dots was measured as a function of temperature at 1 Hz, as well as the local frequency dependence exponent, $\gamma$.  We find that the data can be described reasonably well with our theory using as few as two discrete TLFs,
\begin{equation}\label{eq:TLFsum}
S \!\! = \!\! \int_{0}^{\infty}\!\!\!\!\!\!\!\! dT_{sb} \sum_{i=1}^2 \frac{\Delta_i^2\sqrt{8 m A k_B} e^{-\frac{\left(T_{sb}-T\right)^2}{\frac{6 \hbar^2 T}{\pi m A k_B}}} \text{sech}\!\! \left[\frac{E_i}{k_B T_{sb}} + \ln (\omega\tau_i)\!\right]}{\hbar \omega \sqrt{3 T} \left(1+\text{erf}\left(\sqrt{\frac{\pi m A k_B T}{6\hbar^2}}\right)\right)}
\end{equation}
by fitting over the switching times ($\tau_i$), activation energies ($E_i$), and fluctuator strengths ($\Delta_i^2$), as well as a common 2D sub-bath area ($A$), where we have made the physically reasonable assumption that the heat capacity of the thermal bath is dominated by the electronic contribution and used Eq.~\eqref{eq:sigmasb2}. The objective function simultaneously minimizes net deviations from the moving average of the noisy $S$ and $\gamma$ data with equal weighting, and the minimization is carried out via a local gradient search using the \textit{fmincon} function in Matlab. The fitting parameters are constrained to lie within $0-10$ s for $\tau$, $0-100$ meV for $E_i$, $0-10^4$ meV$^2$ for $\Delta_i^2$, and $2\mathrm{nm}-100\mu\mathrm{m}$ for $\sqrt{A/\pi}$ when it is finite. In the infinite sub-bath case, $A\rightarrow \infty$, the temperature distribution corresponds to a delta function.  The optimization is not guaranteed to find the global minimum, though we perform each optimization over 30 times for each quantum dot with different initial values so as to sample over different local minima.

\begin{figure}[htb]
  \centering
  \includegraphics[width=0.8\columnwidth]{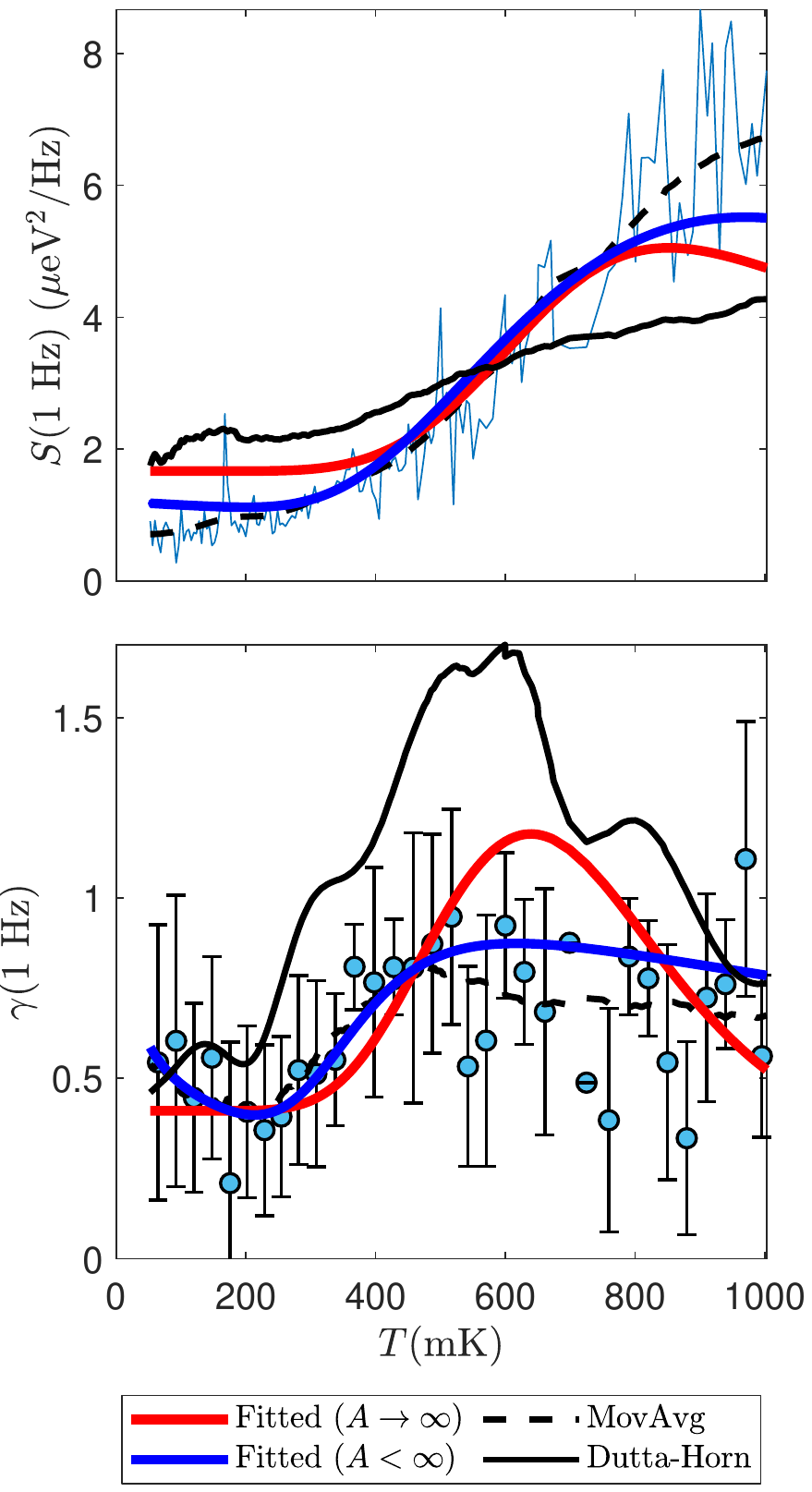}
  \caption{Data from Supplemental Material \cite{supp}, Figs.~2(b) and (d) of Ref.~\cite{Connors_2019} (Device 1, QD R1, Right), along with fits using only two fluctuators. The fitting parameters are given in Table \ref{table:params} and a sensitivity analysis is included in the Supplemental Material  \cite{supp}.}\label{fig:Sandgamma}
\end{figure}
In Fig.~\ref{fig:Sandgamma} we show an example of the results of the fitting with and without taking a microscopic sub-bath area. (Fits to the complete data set are included in the Supplemental Material \cite{supp}.)  Even the fit using an infinite thermal sub-bath appears better than the standard Dutta-Horn results, although this happens because we are just fitting $S$ (a function of $T$ and $\omega$) over a cut at constant $\omega$ while also fitting the derivative $\partial_{\omega} S$ perpendicular to the cut. As we showed at the outset, incorporating a microscopic sub-bath area in principle allows for a $1/\omega$-type frequency dependence over the whole plane.  Even restricting ourselves to the data in hand, there is a noticeable improvement in the fit when including the effects of a microscopic sub-bath. It is interesting that the sub-bath sizes that emerge from the fit are quite consistent across different dots and correspond to disks with radii of about 100 nm, which is physically reasonable for the devices being fitted which have top gate feature sizes ranging from roughly 70 nm to 200 nm \cite{Connors_2019}. Note that these small radii correspond to significantly large temperature fluctuations; for the device corresponding to Fig.~\ref{fig:Sandgamma} and Table \ref{table:params}, for example, the sub-bath radius of $\sim 80$ nm that gives the best fit corresponds to a standard deviation $\sigma_{sb} \sim 330$ mK at a mean temperature of $T \sim 500$ mK.
\begin{table}[]
\begin{tabular}{c|c|c|c|}
\cline{2-4}
                                                & Parameter      & Fluctuator 1   & Fluctuator 2  \\ \hline
\multicolumn{1}{|c|}{\multirow{3}{*}{$A \rightarrow \infty$}}    & $\tau$ (ms)     & 8.397         & 80.803         \\ \cline{2-4}
\multicolumn{1}{|c|}{}                          & $E$ (meV)       & 0.216         & 5.414$\times10^{-11}$ \\ \cline{2-4}
\multicolumn{1}{|c|}{}                          & $\Delta_i^2$ (meV$^2$)   & 10.624          & 6.495              \\ \hline
\multicolumn{1}{|c|}{\multirow{4}{*}{$A < \infty$}} & $\tau$ (ms)       & 1.423$\times10^{-5}$  & 57.104              \\ \cline{2-4}
\multicolumn{1}{|c|}{}                          & $E$ (meV)         & 1.478       & 3.128$\times10^{-3}$        \\ \cline{2-4}
\multicolumn{1}{|c|}{}                          & $\Delta_i^2$ (meV$^2$)     & 80.580      & 4.807           \\ \cline{2-4}
\multicolumn{1}{|c|}{}                          & $\sqrt{A/\pi}$ (nm) & \multicolumn{2}{c|}{80.488} \\ \hline
\end{tabular}
\caption{Fitting parameter values for Fig.~\ref{fig:Sandgamma}.}\label{table:params}
\end{table}
The data can be fit more closely with three or four TLFs (see Supplemental Material \cite{supp}), but given how much variance the data displays, it does not make sense to strive for too much precision in fitting the average.

One ramification of having only a few relevant TLFs would be that increasing temperature may not be as deleterious to coherence as it is for typical $1/f$ noise, as suggested in Ref.~\cite{Petit_2020}.  If it is furthermore true that these TLFs are indeed coupled to a microscopic sub-bath with appreciable temperature fluctuations, it could have some other surprising but testable ramifications.  For instance, it is natural to wonder what happens if the sub-bath is small enough that the effective temperature distribution has a long tail leading to fluctuations larger than the mean temperature. Indeed, from \eqref{eq:sigmasb2}, this is always the case at small enough mean temperature, $T < 3\hbar^2/\pi m A k_B$. For a simple planar bath where there is no other relevant length scale, the sub-bath area should go like the square of the distance, $d$, between the TLF and the bath, so the critical distance at which $\sigma_{sb} \sim T$ is $d_c \sim \hbar/\sqrt{m k_B T}$, which is around 200nm for $T\sim 100$mK in Si. The dependence of the low-frequency PSD amplitude on the distance goes like $\sim d \exp\left(-d^2 k_B T m/\hbar^2 \right)$ [cf.~Eqs.~\eqref{eq:Sapprox} and \eqref{eq:sigmasb2}], which diminishes linearly with decreasing distance below $d_c$ before saturating at the Lorentzian floor.  Thus, one has the counterintuitive possibility of suppressing low-frequency noise by bringing the thermal electronic bath (presumably the capacitively coupled surrounding 2DEG, or the metal gates) in \emph{closer} contact with the TLFs (perhaps charged defects at the oxide interface, or near the semiconductor surface).

In conclusion, we have shown how a $1/f$ noise power spectral density (PSD) with nonlinear temperature dependence, often modeled as arising from a broad distribution of two-level fluctuators (TLFs) via the Dutta-Horn relation, can in fact emerge from even one or two TLFs when coupled to a microscopic thermal sub-bath due to effective temperature fluctuations.  If a broad distribution of TLFs is coupled to such a bath, the strict connection between local frequency and temperature scalings enforced by the Dutta-Horn relation is relaxed.  Finally, we noted that recent experimental measurements of both the local frequency scaling and a nonlinear temperature dependence in silicon quantum dots can be reasonably explained as arising from as few as two TLFs.

\begin{acknowledgments}
The authors thank Elliot Connors for providing the data sets measured in Ref.~\cite{Connors_2019}.  S.A. and S.D.S. acknowledge support by the Laboratory for Physical Sciences. J.P.K. acknowledges support by the Army Research Office (ARO) under Grant No. W911NF-17-1-0287.
\end{acknowledgments}



\section*{Supplementary material}

# Supplemental Material:
# Microscopic bath effects on noise spectra in semiconductor quantum dots

Seongjin Ahn,[1] S. Das Sarma,[1] and J. P. Kestner[2, *]

[1] Condensed Matter Theory Center and Joint Quantum Institute,
Department of Physics, University of Maryland College Park, MD 20742, USA
[2] Department of Physics, University of Maryland Baltimore County, Baltimore, MD 21250, USA

## COST FUNCTION AND SENSITIVITY

The cost function used in the fitting is

$$J = (wJ_S)^2 + J_\gamma^2 \tag{S1}$$

where

$$J_S = \sqrt{\frac{1}{N}\sum_{i=1}^{N}(S_{\text{fit}}(T_i) - S_{\text{data}}(T_i))^2} \tag{S2}$$

and

$$J_\gamma = \sqrt{\frac{1}{N}\sum_{i=1}^{N}(\gamma_{\text{fit}}(T_i) - \gamma_{\text{data}}(T_i))^2} \tag{S3}$$

are the standard deviations of the fit curves for $S$ and $\gamma$ with respect to the measured values across the set of $N$ temperature data points. The datasets $S_{\text{data}}$ and $\gamma_{\text{data}}$ are equally weighted through the weighting factor $w$, accounting for the relative magnitude of $S$ to $\gamma$. In Fig. S1 we show how sensitive the cost function is to multiplicative perturbations of the optimal fit parameters $\tau_{1,2}$ $E_{1,2}$, $\Delta_{1,2}^2$, and $A$ reported in Table I of the main text. Although plotting multiplicative perturbations allows a unified, dimensionless representation of the sensitivity to parameters of different units, one should also note one shortcoming. From the plot it looks like the cost function is completely insensitive to $E_2$ in the $A \to \infty$ case and $\tau_1$ in the $A < \infty$ case, but that is just due to the fact that their optimal values are nearly zero, so a perturbation of a few percent is negligible. The cost function does of course depend on the values of those parameters, and we can give a better sense of the sensitivity by noting that to cause a $\sim 10\%$ change in the cost function requires a perturbation on the order of $\sim 1\mu$eV in $E_2$ for $A \to \infty$ and on the order of $\sim 10$ns in $\tau_1$ for $A < \infty$. So the cost function is well-behaved in the region around the minimum we used.

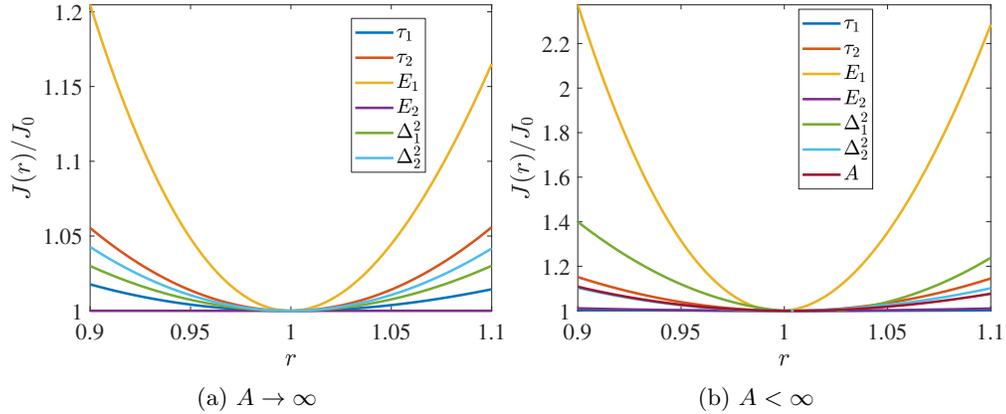

FIG. S1: The cost function plotted as a function of one fitting parameter with the rest being fixed at the optimal values. $r$ is the ratio of the varying fitting parameter relative to its optimal value and $J_0$ is the cost function at the optimal fit parameters reported in Table I of the main text.

**FITTING RESULTS**

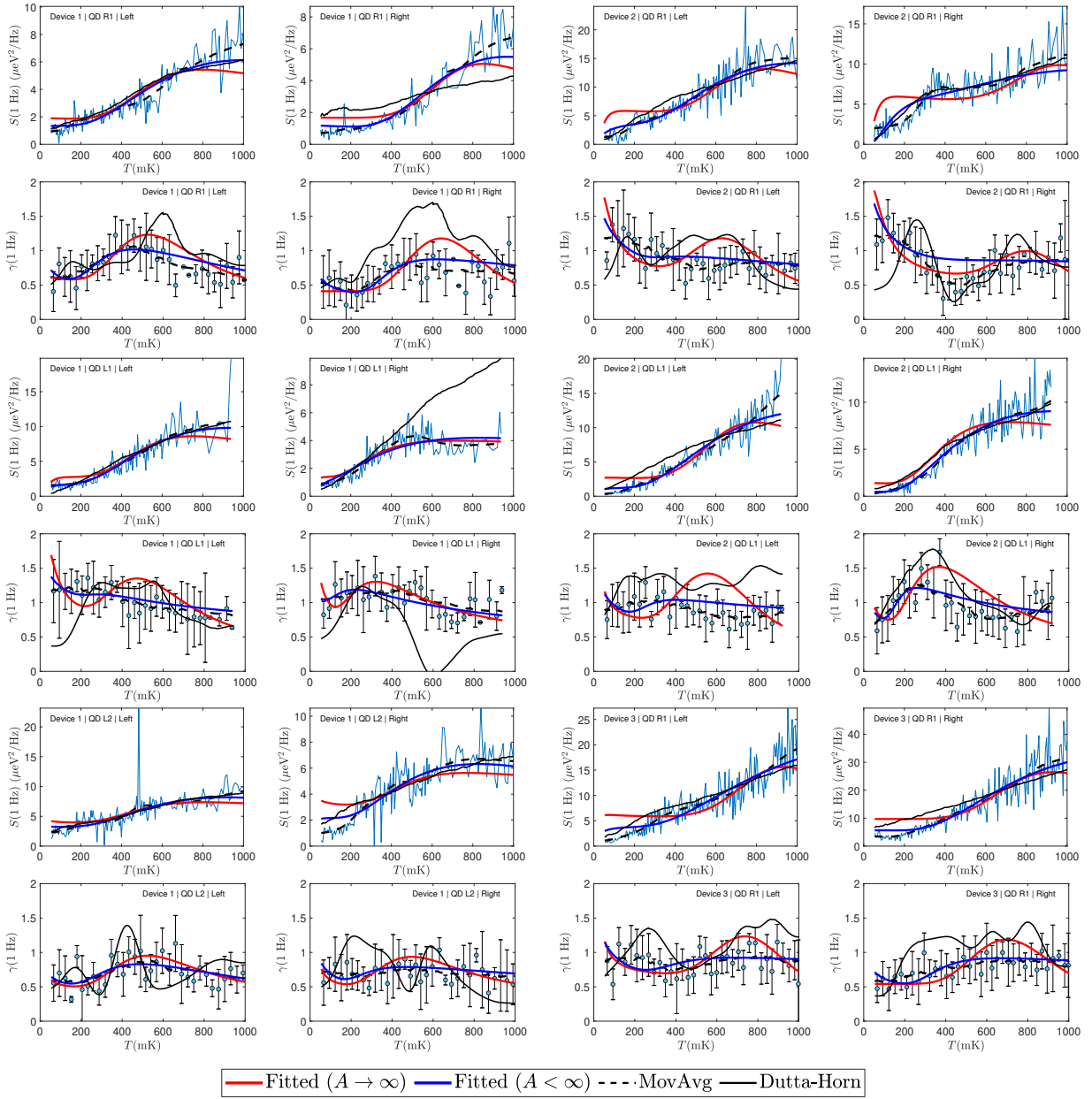

FIG. S2: Fitting results using two fluctuators for $S$ and $\gamma$ along with the DH lines and measured data from Supplementary Fig. 2 of Ref. [1]

|  | | Device 1 QD R1 Left | | Device 1 QD R1 Right | | Device 2 QD R1 Left | | Device 2 QD R1 Right | |
|---|---|---|---|---|---|---|---|---|---|
|  | Parameters | Fluctuator1 | Fluctuator2 | Fluctuator1 | Fluctuator2 | Fluctuator1 | Fluctuator2 | Fluctuator1 | Fluctuator2 |
| $A \to \infty$ | $\tau$ (ms) | 22.561 | 99.471 | 8.397 | 80.803 | 7.874 | 96.097 | 3.033 | 87.777 |
|  | $E$ (meV) | 0.134 | $3.120 \times 10^{-4}$ | 0.216 | $5.415 \times 10^{-11}$ | 0.215 | $6.770 \times 10^{-3}$ | 0.331 | $8.598 \times 10^{-3}$ |
|  | $\Delta^2$ (meV$^2$) | 11.256 | 6.464 | 10.624 | 6.495 | 24.593 | 18.343 | 14.401 | 18.651 |
| $A < \infty$ | $\tau$ (ms) | 1.191 | 69.130 | $1.423 \times 10^{-5}$ | 57.104 | $3.491 \times 10^{-6}$ | 43.513 | 8.072e-07 | 0.264 |
|  | $E$ (meV) | 0.409 | $3.320 \times 10^{-3}$ | 1.478 | $3.128 \times 10^{-3}$ | 1.808 | 0.034 | 2.891 | 0.349 |
|  | $\Delta^2$ (meV$^2$) | 32.726 | 5.127 | 80.580 | 4.807 | 243.612 | 15.701 | 163.006 | 67.286 |
|  | $\sqrt{A/\pi}$ (nm) | 78.708 | | 80.488 | | 61.636 | | 44.761 | |
|  | | Device 1 QD L1 Left | | Device 1 QD L1 Right | | Device 2 QD L1 Left | | Device 2 QD L1 Right | |
| $A \to \infty$ | $\tau$ (ms) | 105.924 | 24.400 | 102.331 | 66.145 | 10.921 | 111.491 | 42.183 | 103.979 |
|  | $E$ (meV) | $5.807 \times 10^{-3}$ | 0.122 | $3.401 \times 10^{-3}$ | 0.055 | 0.184 | $2.374 \times 10^{-3}$ | 0.082 | $1.589 \times 10^{-3}$ |
|  | $\Delta^2$ (meV$^2$) | 8.635 | 18.830 | 4.347 | 8.489 | 25.705 | 8.591 | 20.706 | 4.420 |
| $A < \infty$ | $\tau$ (ms) | $1.973 \times 10^{-4}$ | 101.081 | $1.758 \times 10^{-6}$ | 0.734 | $6.083 \times 10^{-6}$ | 66.274 | 0.014 | 70.068 |
|  | $E$ (meV) | 1.310 | 0.012 | 0.670 | 0.571 | 2.299 | 0.017 | 0.957 | $7.316 \times 10^{-3}$ |
|  | $\Delta^2$ (meV$^2$) | 144.948 | 6.693 | 37.162 | 36.739 | 292.071 | 4.847 | 122.575 | 1.792 |
|  | $\sqrt{A/\pi}$ (nm) | 59.059 | | 36.461 | | 42.228 | | 48.097 | |
|  | | Device 1 QD L2 Left | | Device 1 QD L2 Right | | Device 3 QD R1 Left | | Device 3 QD R1 Right | |
| $A \to \infty$ | $\tau$ (ms) | 86.444 | 31.877 | 34.965 | 83.889 | 105.719 | 3.696 | 8.248 | 96.978 |
|  | $E$ (meV) | $8.874 \times 10^{-4}$ | 0.112 | 0.102 | $1.867 \times 10^{-3}$ | $2.664 \times 10^{-3}$ | 0.304 | 0.237 | $6.162 \times 10^{-13}$ |
|  | $\Delta^2$ (meV$^2$) | 14.350 | 11.006 | 8.232 | 11.301 | 19.333 | 30.988 | 52.620 | 34.289 |
| $A < \infty$ | $\tau$ (ms) | 0.626 | 76.580 | $9.292 \times 10^{-5}$ | 43.877 | $1.000 \times 10^{-5}$ | 34.991 | $6.307 \times 10^{-6}$ | 66.718 |
|  | $E$ (meV) | 0.432 | $1.838 \times 10^{-3}$ | 1.306 | 0.013 | 3.793 | 0.052 | 2.707 | $6.134 \times 10^{-3}$ |
|  | $\Delta^2$ (meV$^2$) | 37.513 | 12.058 | 91.592 | 9.500 | 480.655 | 17.774 | 704.864 | 22.023 |
|  | $\sqrt{A/\pi}$ (nm) | 81.262 | | 56.113 | | 27.471 | | 39.858 | |

TABLE S1: Fitting parameter values for Fig. S2

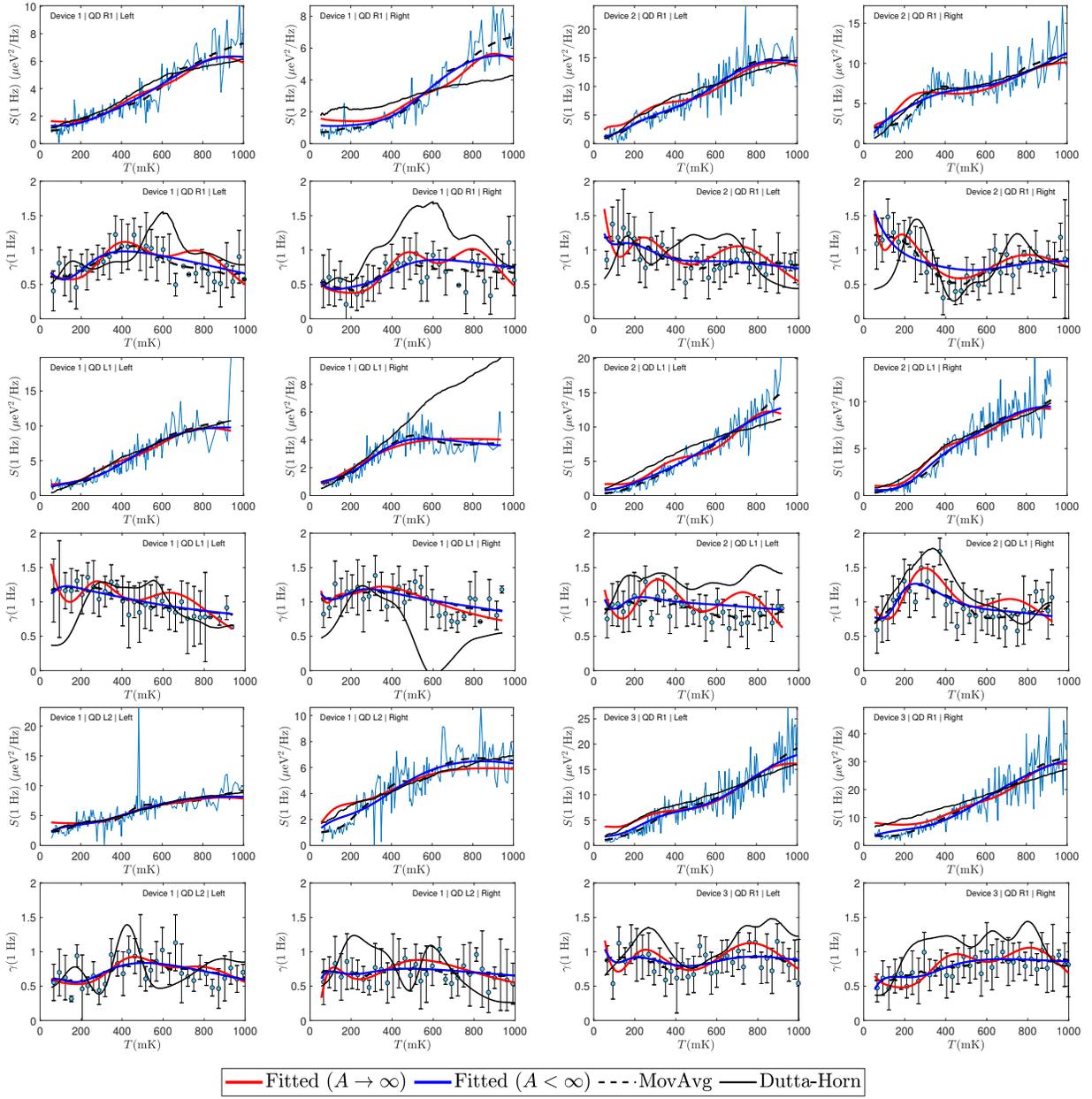

FIG. S3: Fitting results using three fluctuators for $S$ and $\gamma$ along with the DH lines and measured data from Supplementary Fig. 2 of Ref. [1]

|  | Parameters | \multicolumn{3}{c}{Device 1 QD R1 Left} | | | \multicolumn{3}{c}{Device 1 QD R1 Right} | | |
|---|---|---|---|---|---|---|---|
|  |  | Fluctuator1 | Fluctuator2 | Fluctuator3 | Fluctuator1 | Fluctuator2 | Fluctuator3 |
| $A \to \infty$ | $\tau$ (ms) | 1.414 | 15.557 | 94.933 | 71.026 | 0.747 | 4.466 |
|  | $E$ (meV) | 0.371 | 0.115 | $7.108 \times 10^{-4}$ | $1.223 \times 10^{-3}$ | 0.426 | 0.182 |
|  | $\Delta^2$ (meV$^2$) | 10.141 | 6.615 | 5.521 | 5.695 | 10.938 | 4.809 |
| $A < \infty$ | $\tau$ (ms) | $5.863 \times 10^{-6}$ | 13.803 | 65.589 | $1.037 \times 10^{-5}$ | 2.617 | 43.476 |
|  | $E$ (meV) | 1.496 | 0.142 | $2.409 \times 10^{-3}$ | 1.395 | 0.214 | $3.881 \times 10^{-3}$ |
|  | $\Delta^2$ (meV$^2$) | 35.275 | 10.835 | 4.988 | 60.890 | 4.014 | 5.136 |
|  | $\sqrt{A/\pi}$ (nm) | | 124.730 | | | 107.047 | |
|  |  | \multicolumn{3}{c}{Device 2 QD R1 Left} | | | \multicolumn{3}{c}{Device 2 QD R1 Right} | | |
| $A \to \infty$ | $\tau$ (ms) | 5.975 | 31.285 | 70.631 | 35.268 | 65.374 | 10.534 |
|  | $E$ (meV) | 0.256 | 0.056 | $6.637 \times 10^{-3}$ | 0.041 | $6.543 \times 10^{-3}$ | 0.246 |
|  | $\Delta^2$ (meV$^2$) | 27.081 | 14.381 | 9.409 | 13.653 | 7.827 | 17.492 |
| $A < \infty$ | $\tau$ (ms) | $5.420 \times 10^{-6}$ | 3.206 | 26.804 | $2.387 \times 10^{-6}$ | $2.591 \times 10^{-5}$ | 17.757 |
|  | $E$ (meV) | 1.487 | 0.157 | 0.021 | 0.990 | 1.935 | 0.049 |
|  | $\Delta^2$ (meV$^2$) | 161.450 | 27.991 | 7.027 | 66.674 | 143.223 | 20.335 |
|  | $\sqrt{A/\pi}$ (nm) | | 106.034 | | | 87.483 | |
|  |  | \multicolumn{3}{c}{Device 1 QD L1 Left} | | | \multicolumn{3}{c}{Device 1 QD L1 Right} | | |
| $A \to \infty$ | $\tau$ (ms) | 21.54 | 98.967 | 8.652 | 39.822 | 67.294 | 82.135 |
|  | $E$ (meV) | 0.074 | $5.084 \times 10^{-3}$ | 0.213 | 0.028 | 0.060 | $3.757 \times 10^{-3}$ |
|  | $\Delta^2$ (meV$^2$) | 9.355 | 5.349 | 19.324 | 2.176 | 8.931 | 2.887 |
| $A < \infty$ | $\tau$ (ms) | $5.708 \times 10^{-6}$ | 0.018 | 30.506 | $4.003 \times 10^{-6}$ | 75.868 | 90.409 |
|  | $E$ (meV) | 1.562 | 0.106 | 0.072 | 0.758 | 0.059 | $3.879 \times 10^{-3}$ |
|  | $\Delta^2$ (meV$^2$) | 155.949 | 16.180 | 10.660 | 19.242 | 7.676 | 3.612 |
|  | $\sqrt{A/\pi}$ (nm) | | 68.738 | | | 144.893 | |
|  |  | \multicolumn{3}{c}{Device 2 QD L1 Left} | | | \multicolumn{3}{c}{Device 2 QD L1 Right} | | |
| $A \to \infty$ | $\tau$ (ms) | 1.352 | 19.094 | 89.447 | 105.672 | 18.485 | 5.352 |
|  | $E$ (meV) | 0.360 | 0.087 | $3.446 \times 10^{-3}$ | $1.406 \times 10^{-3}$ | 0.087 | 0.266 |
|  | $\Delta^2$ (meV$^2$) | 25.647 | 12.460 | 5.281 | 3.267 | 13.467 | 17.579 |
| $A < \infty$ | $\tau$ (ms) | $2.780 \times 10^{-6}$ | 0.471 | 10.305 | $2.190 \times 10^{-5}$ | 0.357 | 114.402 |
|  | $E$ (meV) | 2.079 | 0.360 | 0.039 | 1.394 | 0.289 | $3.226 \times 10^{-4}$ |
|  | $\Delta^2$ (meV$^2$) | 208.734 | 35.862 | 5.624 | 74.733 | 34.818 | 1.892 |
|  | $\sqrt{A/\pi}$ (nm) | | 69.218 | | | 125.596 | |
|  |  | \multicolumn{3}{c}{Device 1 QD L2 Left} | | | \multicolumn{3}{c}{Device 1 QD L2 Right} | | |
| $A \to \infty$ | $\tau$ (ms) | 3.934 | 9.550 | 90.957 | 43.486 | 74.819 | 1.524 |
|  | $E$ (meV) | 0.306 | 0.142 | $7.093 \times 10^{-4}$ | 0.100 | 0.015 | $2.346 \times 10^{-8}$ |
|  | $\Delta^2$ (meV$^2$) | 8.741 | 8.132 | 13.192 | 9.369 | 5.370 | 242.536 |
| $A < \infty$ | $\tau$ (ms) | 2.434 | 4.533 | 37.055 | $4.152 \times 10^{-5}$ | 2.603 | 16.395 |
|  | $E$ (meV) | 0.327 | 0.070 | $5.748 \times 10^{-6}$ | 1.362 | $2.800 \times 10^{-6}$ | 0.072 |
|  | $\Delta^2$ (meV$^2$) | 35.279 | 12.752 | 10.223 | 85.572 | 79.605 | 8.822 |
|  | $\sqrt{A/\pi}$ (nm) | | 84.888 | | | 63.148 | |
|  |  | \multicolumn{3}{c}{Device 3 QD R1 Left} | | | \multicolumn{3}{c}{Device 3 QD R1 Right} | | |
| $A \to \infty$ | $\tau$ (ms) | 68.903 | 36.885 | 5.639 | 11.055 | 1.820 | 81.042 |
|  | $E$ (meV) | $4.674 \times 10^{-3}$ | 0.052 | 0.284 | 0.135 | 0.378 | $1.444 \times 10^{-3}$ |
|  | $\Delta^2$ (meV$^2$) | 11.713 | 11.863 | 32.297 | 21.655 | 56.113 | 27.313 |
| $A < \infty$ | $\tau$ (ms) | $2.237 \times 10^{-4}$ | 0.069 | 41.893 | $1.826 \times 10^{-5}$ | 20.921 | 25.238 |
|  | $E$ (meV) | 1.384 | 0.336 | 0.012 | 2.348 | 0.097 | $7.445 \times 10^{-5}$ |
|  | $\Delta^2$ (meV$^2$) | 184.023 | 48.086 | 8.369 | 621.410 | 19.321 | 27.002 |
|  | $\sqrt{A/\pi}$ (nm) | | 102.364 | | | 46.331 | |

TABLE S2: Fitting parameter values for Fig. S3

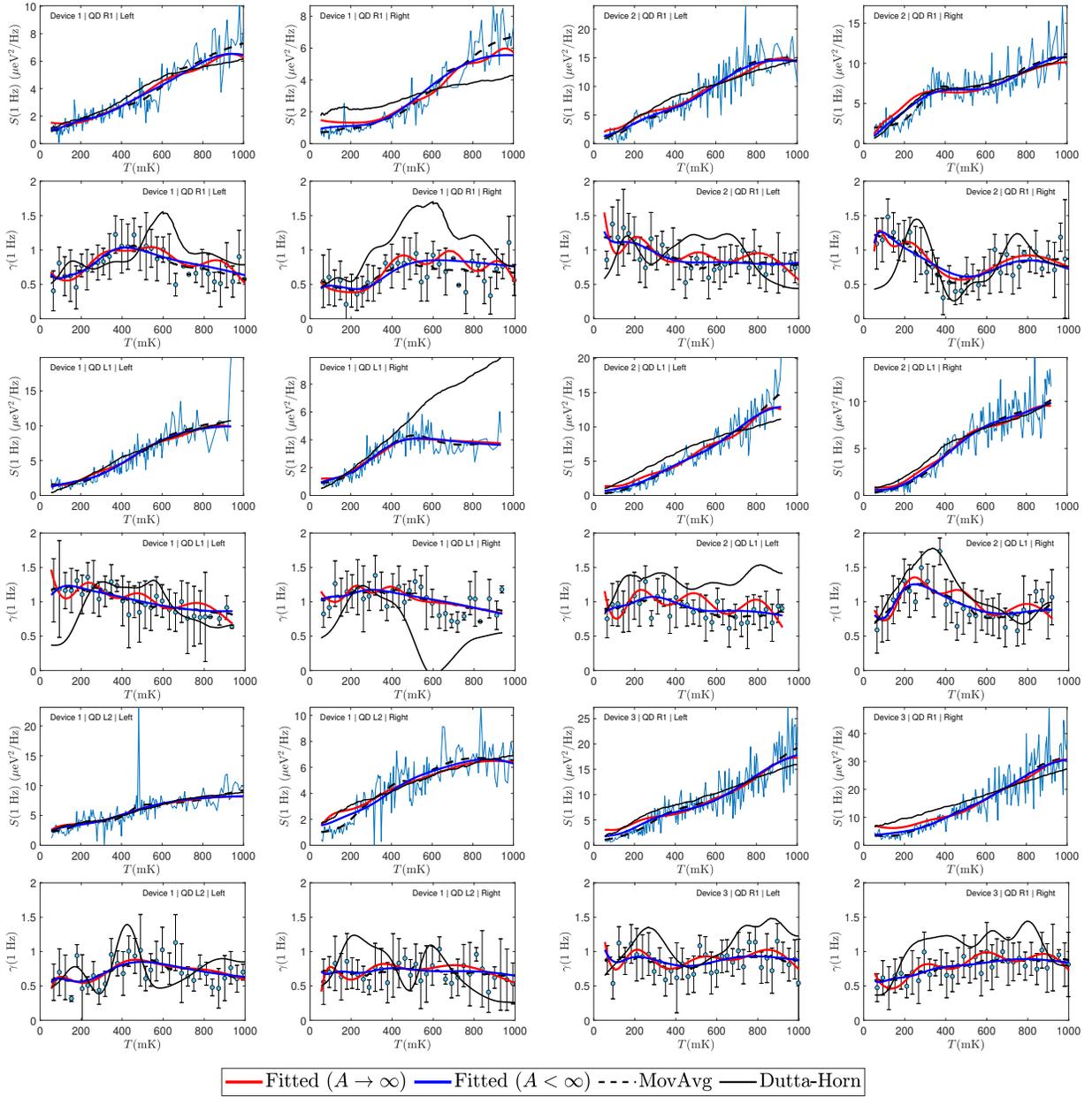

FIG. S4: Fitting results using four fluctuators for $S$ and $\gamma$ along with the DH lines and measured data from Supplementary Fig. 2 of Ref. [1]

| | Parameters | \multicolumn{4}{c}{Device 1 QD R1 Left} | | | | \multicolumn{4}{c}{Device 1 QD R1 Right} | | | |
|---|---|---|---|---|---|---|---|---|---|
| | | Fluctuator1 | Fluctuator2 | Fluctuator3 | Fluctuator4 | Fluctuator1 | Fluctuator2 | Fluctuator3 | Fluctuator4 |
| $A \to \infty$ | $\tau$ (ms) | 2.422 | 0.151 | 87.038 | 23.495 | 0.119 | 2.967 | 71.166 | $7.938 \times 10^{-3}$ |
| | $E$ (meV) | 0.246 | 0.577 | $1.218 \times 10^{-3}$ | 0.081 | 0.473 | 0.186 | $1.368 \times 10^{-3}$ | 0.837 |
| | $\Delta^2$ (meV$^2$) | 7.005 | 9.688 | 5.081 | 3.847 | 8.236 | 3.873 | 5.263 | 10.129 |
| $A < \infty$ | $\tau$ (ms) | $2.080 \times 10^{-5}$ | $3.273 \times 10^{-5}$ | 4.411 | 49.626 | $1.815 \times 10^{-5}$ | 0.113 | 40.023 | 210.174 |
| | $E$ (meV) | 1.385 | 0.318 | 0.194 | $3.263 \times 10^{-3}$ | 1.451 | $6.253 \times 10^{-5}$ | 0.031 | 4.107 |
| | $\Delta^2$ (meV$^2$) | 33.195 | 11.613 | 13.459 | 4.185 | 78.615 | 1330.979 | 2.114 | 160.203 |
| | $\sqrt{A/\pi}$ (nm) | \multicolumn{4}{c}{162.327} | | | | \multicolumn{4}{c}{82.203} | | | |
| | | \multicolumn{4}{c}{Device 2 QD R1 Left} | | | | \multicolumn{4}{c}{Device 2 QD R1 Right} | | | |
| $A \to \infty$ | $\tau$ (ms) | 2.371 | 1.846 | 29.716 | 69.735 | 12.739 | 58.062 | 25.659 | 0.990 |
| | $E$ (meV) | 0.220 | 0.366 | 0.049 | $6.351 \times 10^{-3}$ | 0.233 | 0.014 | 0.055 | $8.509 \times 10^{-5}$ |
| | $\Delta^2$ (meV$^2$) | 13.221 | 28.097 | 11.527 | 7.575 | 18.292 | 7.540 | 11.839 | 147.797 |
| $A < \infty$ | $\tau$ (ms) | $1.293 \times 10^{-5}$ | 6.648 | 43.699 | 87.858 | 0.196 | 0.313 | 7.302 | 27.798 |
| | $E$ (meV) | 1.288 | 0.117 | 0.014 | 0.697 | 0.588 | 0.229 | $1.212 \times 10^{-6}$ | 0.034 |
| | $\Delta^2$ (meV$^2$) | 122.307 | 22.111 | 6.811 | 1504.409 | 40.849 | 19.822 | 12.390 | 13.390 |
| | $\sqrt{A/\pi}$ (nm) | \multicolumn{4}{c}{124.438} | | | | \multicolumn{4}{c}{165.239} | | | |
| | | \multicolumn{4}{c}{Device 1 QD L1 Left} | | | | \multicolumn{4}{c}{Device 1 QD L1 Right} | | | |
| $A \to \infty$ | $\tau$ (ms) | 2.802 | 23.609 | 2.051 | 105.68 | 6.418 | 136.169 | 2.272 | 48.665 |
| | $E$ (meV) | 0.320 | 0.061 | 0.217 | $4.242 \times 10^{-3}$ | 0.080 | $1.420 \times 10^{-3}$ | 0.170 | 0.094 |
| | $\Delta^2$ (meV$^2$) | 18.937 | 6.791 | 10.450 | 4.648 | 2.616 | 3.823 | 3.618 | 6.607 |
| $A < \infty$ | $\tau$ (ms) | $9.549 \times 10^{-5}$ | $1.843 \times 10^{-4}$ | 0.531 | 22.853 | $3.506 \times 10^{-5}$ | 0.711 | 28.180 | 62.105 |
| | $E$ (meV) | 0.164 | 1.096 | 1.110 | 0.075 | 0.618 | 0.024 | 0.027 | 0.072 |
| | $\Delta^2$ (meV$^2$) | 27.067 | 109.064 | 65.545 | 10.791 | 15.181 | 6.861 | 3.927 | 9.372 |
| | $\sqrt{A/\pi}$ (nm) | \multicolumn{4}{c}{81.839} | | | | \multicolumn{4}{c}{168.886} | | | |
| | | \multicolumn{4}{c}{Device 2 QD L1 Left} | | | | \multicolumn{4}{c}{Device 2 QD L1 Right} | | | |
| $A \to \infty$ | $\tau$ (ms) | 0.260 | 23.27 | 2.816 | 74.646 | 96.990 | 17.379 | 2.710 | 1.703 |
| | $E$ (meV) | 0.497 | 0.058 | 0.210 | $4.174 \times 10^{-3}$ | $1.712 \times 10^{-3}$ | 0.070 | 0.202 | 0.366 |
| | $\Delta^2$ (meV$^2$) | 26.209 | 6.428 | 13.824 | 4.269 | 2.745 | 6.817 | 12.724 | 18.908 |
| $A < \infty$ | $\tau$ (ms) | $5.273 \times 10^{-6}$ | 0.638 | 16.286 | 41.044 | $2.561 \times 10^{-5}$ | $2.590 \times 10^{-3}$ | 4.156 | 88.299 |
| | $E$ (meV) | 1.488 | 0.263 | 0.053 | $5.298 \times 10^{-3}$ | 0.946 | 1.102 | 0.147 | $1.592 \times 10^{-3}$ |
| | $\Delta^2$ (meV$^2$) | 116.015 | 25.533 | 4.036 | 2.515 | 42.629 | 55.535 | 15.554 | 2.016 |
| | $\sqrt{A/\pi}$ (nm) | \multicolumn{4}{c}{142.976} | | | | \multicolumn{4}{c}{159.668} | | | |
| | | \multicolumn{4}{c}{Device 1 QD L2 Left} | | | | \multicolumn{4}{c}{Device 1 QD L2 Right} | | | |
| $A \to \infty$ | $\tau$ (ms) | 110.546 | 3.693 | 10.571 | 18.448 | 17.298 | 64.598 | 15.684 | 1.688 |
| | $E$ (meV) | $9.377 \times 10^{-3}$ | $3.997 \times 10^{-12}$ | 0.246 | 0.116 | 0.092 | 0.014 | 0.194 | $2.136 \times 10^{-5}$ |
| | $\Delta^2$ (meV$^2$) | 5.060 | 133.640 | 8.315 | 8.427 | 5.676 | 4.405 | 9.727 | 177.361 |
| $A < \infty$ | $\tau$ (ms) | 8.544 | 15.725 | 23.393 | 71.422 | $2.550 \times 10^{-5}$ | 0.016 | 3.403 | 42.903 |
| | $E$ (meV) | 1.940e-07 | 0.137 | 0.242 | 0.012 | 1.363 | 0.475 | 0.095 | $4.729 \times 10^{-3}$ |
| | $\Delta^2$ (meV$^2$) | 46.937 | 13.079 | 8.802 | 5.816 | 38.627 | 26.251 | 7.296 | 6.416 |
| | $\sqrt{A/\pi}$ (nm) | \multicolumn{4}{c}{193.821} | | | | \multicolumn{4}{c}{147.142} | | | |
| | | \multicolumn{4}{c}{Device 3 QD R1 Left} | | | | \multicolumn{4}{c}{Device 3 QD R1 Right} | | | |
| $A \to \infty$ | $\tau$ (ms) | 29.569 | 66.689 | 5.789 | 1.180 | 1.889 | 69.086 | 18.540 | 0.544 |
| | $E$ (meV) | 0.048 | $4.705 \times 10^{-3}$ | 0.191 | 0.426 | 0.270 | $2.393 \times 10^{-3}$ | 0.079 | 0.495 |
| | $\Delta^2$ (meV$^2$) | 9.525 | 9.581 | 12.952 | 33.822 | 28.862 | 23.091 | 12.347 | 57.767 |
| $A < \infty$ | $\tau$ (ms) | $5.361 \times 10^{-4}$ | 0.117 | 0.597 | 42.037 | $2.189 \times 10^{-5}$ | $2.267 \times 10^{-4}$ | 8.345 | 42.375 |
| | $E$ (meV) | 1.285 | 0.308 | 0.262 | 0.012 | 1.544 | 0.775 | 0.086 | $3.492 \times 10^{-3}$ |
| | $\Delta^2$ (meV$^2$) | 171.113 | 44.764 | 0.016 | 8.335 | 282.163 | 111.755 | 12.034 | 16.483 |
| | $\sqrt{A/\pi}$ (nm) | \multicolumn{4}{c}{103.559} | | | | \multicolumn{4}{c}{120.581} | | | |

TABLE S3: Fitting parameter values for Fig. S4

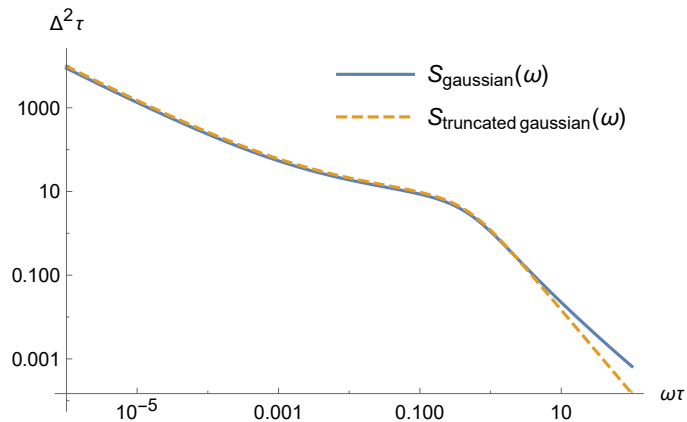

FIG. S5: Comparison of power spectral density vs frequency using a gaussian temperature distribution (Eq. (4) of the main text) and a truncated gaussian (Eq. (S4) of this supplement) for $\frac{E}{k_B T} = 1$, $\frac{\sigma_{sb}}{T} = 0.8$.

## TRUNCATED TEMPERATURE DISTRIBUTION

We now derive Eq. (8) of the main text, starting from Eq. (4), using a truncated gaussian distribution to avoid including negative temperatures.

$$S(\omega) = \frac{2\Delta^2}{\omega \sqrt{\pi/2} \sigma_{sb} \left(1 + \operatorname{erf}\left(\frac{T}{\sqrt{2}\sigma_{sb}}\right)\right)} \int_0^\infty dT_{sb} e^{-\frac{(T_{sb}-T)^2}{2\sigma_{sb}^2}} \operatorname{sech}\left(\frac{E}{k_B}\left(\frac{1}{T_{sb}} - \frac{1}{T_\omega}\right)\right) \quad (S4)$$

$$\approx \frac{2\Delta^2}{\omega} \operatorname{sech}\left(\frac{E}{k_B}\left(\frac{1}{T} - \frac{1}{T_\omega}\right)\right) + \frac{2\Delta^2}{\omega} \frac{\sigma_\omega}{\sqrt{\sigma_{sb}^2 + \sigma_\omega^2}} \frac{1 + \operatorname{erf}\left(\frac{\sigma_{sb}^2 T_\omega + \sigma_\omega^2 T}{\sqrt{2}\sigma_{sb}\sigma_\omega \sqrt{\sigma_{sb}^2 + \sigma_\omega^2}}\right)}{1 + \operatorname{erf}\left(\frac{T}{\sqrt{2}\sigma_{sb}}\right)} e^{-\frac{(T-T_\omega)^2}{2(\sigma_{sb}^2 + \sigma_\omega^2)}}, \quad (S5)$$

assuming well-separated peaks at $T$ and $T_\omega$ as before. In the low-frequency limit $T_\omega \ll T$ and $\sigma_\omega \ll \sigma_{sb}$,

$$S(\omega) \approx \frac{4\tau \Delta^2}{1 + \omega^2 \tau^2} + \frac{2E\Delta^2}{k_B \sigma_{sb} \omega \ln^2 \frac{1}{\omega\tau}} \frac{2 e^{-\frac{(T-T_\omega)^2}{2\sigma_{sb}^2}}}{1 + \operatorname{erf}\left(\frac{T}{\sqrt{2}\sigma_{sb}}\right)}. \quad (S6)$$

Note that the effect on the integral of the negative temperatures in the tail of the untruncated gaussian used in the main text is typically negligible. To illustrate this, we have plotted the power spectral densities resulting from each distribution in Fig. S5. In order to make them at all visually distinguishable, we have had to use an unusually broad distribution with a standard deviation nearly equal to the mean.



\end{document}